%% file: main.tex
\pdfoutput=1
\documentclass[10pt]{article}

\newcommand{\repoURL}{\url{https://github.com/AraoufBh/tokwatt}}
\newcommand{\appURL}{\url{https://araoufbh.github.io/tokwatt/}}
\newcommand{\version}{0.1.0}
\usepackage[a4paper,margin=1in]{geometry}
\usepackage[sc]{mathpazo}
\usepackage[scaled=0.92]{helvet}
\usepackage{microtype}
\usepackage{amsmath}
\usepackage{graphicx}
\usepackage{booktabs}
\usepackage{array}
\usepackage{tabularx}
\usepackage[font=small,labelfont=bf]{caption}
\usepackage[round,authoryear]{natbib}
\usepackage{enumitem}
\usepackage{xcolor}
\usepackage{tikz}
\usetikzlibrary{arrows.meta}
\definecolor{twBlue}{HTML}{2155E8}
\definecolor{twNavy}{HTML}{0B1640}
\definecolor{twGrey}{HTML}{5F676B}
\definecolor{twWash}{HTML}{F2F5FD}
\definecolor{twLine}{HTML}{C9D3EA}
\usepackage{xurl}
\usepackage[colorlinks=true,linkcolor={black!70!blue},citecolor={black!70!blue},urlcolor={black!70!blue}]{hyperref}

\setlist{itemsep=2pt,topsep=4pt}

\definecolor{orcidgreen}{HTML}{A6CE39}
\definecolor{scholarblue}{HTML}{4285F4}
\definecolor{linkedinblue}{HTML}{0A66C2}
\newcommand{\orcidID}[1]{\href{#1}{\begin{tikzpicture}[baseline=-0.55ex]
  \node[circle,fill=orcidgreen,inner sep=0pt,minimum size=2.0ex]{};
  \node[text=white,font=\bfseries,inner sep=0pt]{\fontsize{4.6}{4.6}\selectfont iD};\end{tikzpicture}}}
\newcommand{\scholarID}[1]{\href{#1}{\begin{tikzpicture}[baseline=-0.55ex,y=1ex,x=1ex]
  \fill[scholarblue] (0,0.62) -- (1.0,1.16) -- (2.0,0.62) -- (1.0,0.08) -- cycle;
  \fill[scholarblue] (0.5,0.46) .. controls (1.0,0.22) .. (1.5,0.46) -- (1.5,0.16)
        .. controls (1.0,-0.06) .. (0.5,0.16) -- cycle;\end{tikzpicture}}}
\newcommand{\linkedinID}[1]{\href{#1}{\begin{tikzpicture}[baseline=-0.55ex]
  \node[rounded corners=0.35ex,fill=linkedinblue,inner sep=0pt,minimum size=2.0ex]{};
  \node[text=white,font=\bfseries,inner sep=0pt]{\fontsize{5}{5}\selectfont in};\end{tikzpicture}}}
\newcommand{\AuthorOrcidURL}{https://orcid.org/0009-0003-7116-5080}
\newcommand{\AuthorScholarURL}{https://scholar.google.com/citations?user=N2il6NQAAAAJ\&hl=en\&oi=ao}
\newcommand{\AuthorLinkedinURL}{https://www.linkedin.com/in/abderaouf-bahi-phd-7b934a195/}
\newcommand{\authoricons}{\orcidID{\AuthorOrcidURL}\,\scholarID{\AuthorScholarURL}\,\linkedinID{\AuthorLinkedinURL}}
\newcommand{\file}[1]{\texttt{#1}}
\newcommand{\Wh}{\,Wh}

\input{generated/stats}

\title{\textbf{TokWatt: an open, boundary-annotated database of AI inference energy, from chat answers to agents}}
\author{Abderaouf Bahi\,\authoricons\\[2pt]
LIMA Lab, Faculty of Science and Technology, \\Chadli Bendjedid University, El Tarf 36000, Algeria\\[2pt]
\href{mailto:a.bahi@univ-eltarf.dz}{\texttt{a.bahi@univ-eltarf.dz}}}
\date{}

\hypersetup{pdftitle={TokWatt: an open, boundary-annotated database of AI inference energy, from chat answers to agents},
  pdfauthor={Abderaouf Bahi}}

\begin{document}
\maketitle

\begin{abstract}
\noindent Electricity supply now limits how fast artificial intelligence (AI) can be deployed, yet the public evidence on how much energy one AI task uses is scattered and hard to compare. Published figures for a single chatbot answer differ by two orders of magnitude, mostly because studies count different parts of the system and measure different units of work. We present TokWatt, an open database of \nEst{} published estimates of AI inference energy from \nSrcEst{} sources (\yearMin--\yearMax), covering \nTasks{} tasks from classic machine-learning inferences to chat answers, reasoning, long-document queries, image and video generation and coding agents. Each estimate records its functional unit, five measurement-boundary components, method, reported interval and the location of the value in its source. A reproducible procedure scales estimates to a common server-plus-facility boundary and derives per-task reference values with ranges. The reference values span \oomWord{} orders of magnitude: \refChat\Wh{} for a short chat answer (range \refChatLo--\refChatHi), \refReason\Wh{} for a reasoning answer, \refVideo\Wh{} for a short video clip and \refAgent\Wh{} for one task delegated to a coding agent. Five independent estimates for default production chat systems agree within a factor of \clusterRatio, and the reference values move by at most \sensMaxPct\% under alternative harmonization parameters. A simple identity links personal use to grid load: $E$ watt-hours per person per day across one billion people is an average draw of $E/24$ gigawatts. Data and code are released at \repoURL{}. The calculator is available online at \appURL.
\end{abstract}


% ===========================================================================
\section{Introduction}
\label{sec:intro}

Electricity has become a binding constraint on artificial intelligence. The International Energy Agency (IEA) estimates that data centres used about \ieaTwhA~TWh in 2024, around \ieaShare\% of global electricity consumption, and projects \ieaTwhB~TWh by 2030 in its base case \citep{iea2025energyai}. Gartner expects worldwide data-centre consumption to rise from \gartnerTwhA~TWh in 2025 to \gartnerTwhB~TWh in 2026, with servers optimized for AI accounting for \gartnerAIShare\% of the 2026 total, and to exceed \gartnerTwhC~TWh by 2030 \citep{gartner2026dc}. In the United States, data centres used \lbnlTwh~TWh in 2023, \lbnlShare\% of national consumption, and could use \lbnlTwhLo--\lbnlTwhHi~TWh by 2028, or \lbnlShareLo--\lbnlShareHi\% \citep{shehabi2024lbnl}. These aggregates shape investment, grid planning and regulation, but they do not tell a user, a product team or a policy analyst what one use of AI costs in energy.

Per-task figures are where much of the public debate takes place, and they disagree. An early estimate put a ChatGPT request at about 3\Wh{} \citep{devries2023growing}; a later bottom-up model found about 0.3\Wh{} \citep{you2025chatgpt}; OpenAI's chief executive stated 0.34\Wh{} for an average query \citep{altman2025gentle}; Google measured 0.24\Wh{} for the median Gemini Apps text prompt, or 0.10\Wh{} if only the active accelerators are counted \citep{elsworth2025measuring}. Measurements of open models range from about \mitLo{} to \mitHi\Wh{} per response depending mainly on model size \citep{odonnell2025math}, and modeled estimates for reasoning models reach tens of watt-hours \citep{jegham2025hungry}. Most of this spread is not physics. It comes from two choices that are often left implicit: \emph{what is counted}, the measurement boundary (accelerators only, the whole server, idle capacity, facility overhead, the manufacture of the hardware), and \emph{what is being done}, the functional unit (a short answer, a reasoning chain, a long document, an agent that makes many model calls, an image, a video clip).

Several resources measure or model inference energy, but each works with one method and one boundary. Benchmarks measure the energy of open models on fixed hardware \citep{luccioni2024power,chung2025mlenergy,aienergyscore2025}; models infer the energy of commercial services from throughput and hardware assumptions \citep{jegham2025hungry,oviedo2026energy,rince2025ecologits}; and a few companies disclose first-party figures \citep{elsworth2025measuring,altman2025gentle,mistral2025lca}. What is missing is a place where published estimates are recorded side by side with their boundaries and functional units, so that like can be compared with like, and from which per-task values with explicit uncertainty can be derived reproducibly and updated as new evidence appears.

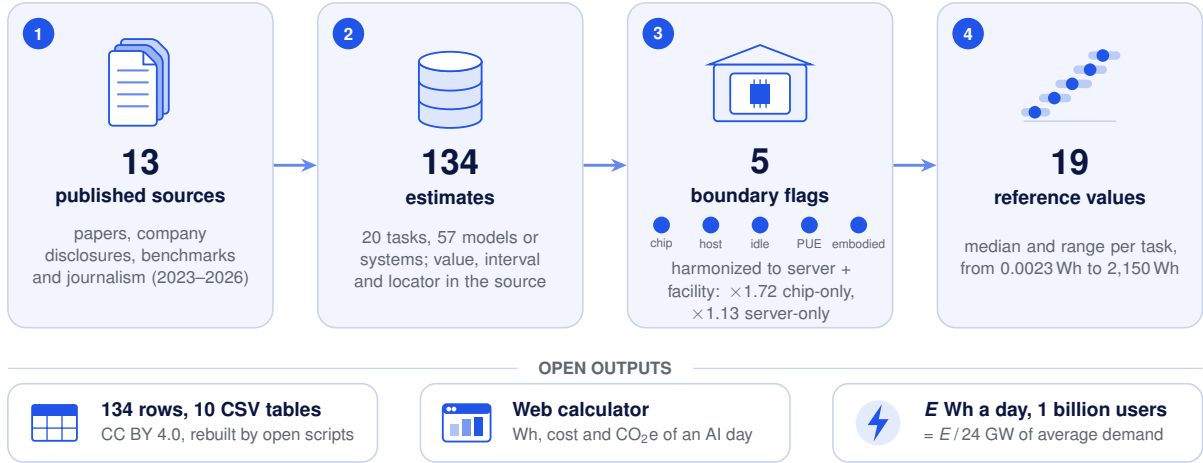
\begin{figure}[t]
  \centering
  \resizebox{\linewidth}{!}{\input{figures/fig_overview}}
  \caption{Overview of TokWatt. Published sources (1) yield estimates recorded with their functional unit, interval and location in the source (2); each estimate carries five boundary flags and is harmonized to one boundary (3); harmonized estimates give a reference value and range per task (4). The release includes the tables, a web calculator and a rule that scales personal use to grid demand.}
  \label{fig:overview}
\end{figure}

\begin{figure}[t]
  \centering
  \includegraphics[width=\linewidth]{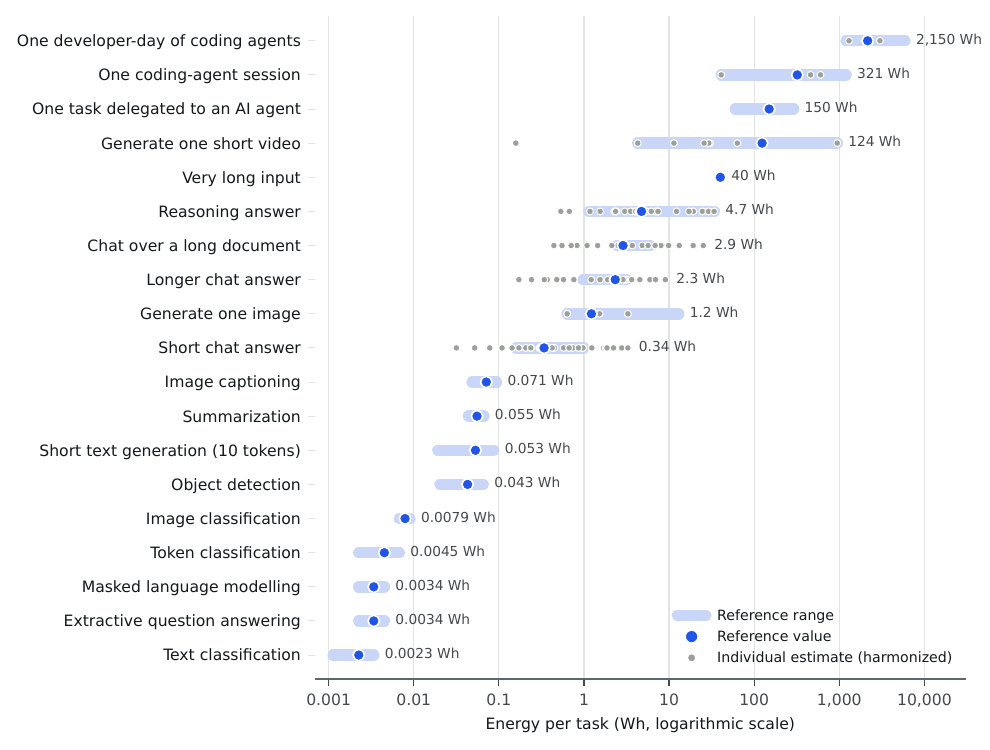}
  \caption{The energy ladder of AI inference. For each task, the dark dot is the TokWatt reference value (median of the task's reference set after harmonization), the band is the reference range, and grey dots are all harmonized estimates for that task, including those outside the reference set. Logarithmic scale; values in Table~\ref{tab:reference}.}
  \label{fig:ladder}
\end{figure}

This paper presents TokWatt (version~\version), an open database and toolkit built for that purpose (Figure~\ref{fig:overview}). Its contributions are:
\begin{enumerate}
  \item a database of \nEst{} published estimates from \nSrcEst{} sources, each recorded with its functional unit, five boundary components, method, interval, token counts where reported, and the location of the value in the source (Sections~\ref{sec:collection} and~\ref{sec:records});
  \item a boundary taxonomy and a harmonization procedure with stated parameters, checked against a paired measurement and tested for sensitivity (Sections~\ref{sec:harmonization} and~\ref{sec:validation});
  \item reference values with ranges for \nTasksRef{} tasks and an analysis of what they imply, from a single answer to population-scale demand (Section~\ref{sec:results} and Figure~\ref{fig:ladder});
  \item open tools: scripts that rebuild and check every derived value, and a calculator that runs in the browser and turns a day of AI use into watt-hours, cost and emissions, scaled up to a billion users (Section~\ref{sec:app}).
\end{enumerate}

% ===========================================================================
\section{Related work}
\label{sec:related}

Concern about the energy of machine learning grew with model size. \citet{strubell2019energy} estimated the energy and carbon of training large language-processing models, \citet{schwartz2020green} argued for reporting efficiency alongside accuracy, and \citet{henderson2020towards} proposed a framework for the systematic reporting of energy and carbon. Calculators \citep{lacoste2019quantifying,lannelongue2021green} and tracking software such as CodeCarbon \citep{courty2026codecarbon} made such accounting practical. Studies of large training runs quantified the effect of hardware, data-centre efficiency and grid location \citep{patterson2021carbon,patterson2022carbon,dodge2022measuring}, and others widened the scope to the full life cycle, including hardware manufacturing \citep{wu2022sustainable,luccioni2023estimating,faiz2024llmcarbon}, and to water \citep{li2025thirsty}. As deployment grew, attention moved to inference: \citet{desislavov2023trends} analysed trends in inference energy, \citet{samsi2023words} benchmarked the energy of large-language-model inference on research GPUs, and \citet{fernandez2025energy} showed how strongly inference energy depends on the workload and on serving optimizations.

Three kinds of per-task evidence now exist. \emph{Measurements} run open models on known hardware: \citet{luccioni2024power} measured ten tasks across task-specific and general-purpose models; the ML.ENERGY Benchmark measures the GPU energy of open models in production-like serving \citep{chung2025mlenergy}; the AI Energy Score rates models by GPU energy in standardized runs on H100 GPUs \citep{aienergyscore2025}; and \citet{delavande2025videoblog} measured open text-to-video models. \emph{Models} estimate the energy of commercial services from public throughput, latency and hardware data: \citet{jegham2025hungry} benchmark dozens of models served through commercial interfaces, including reasoning models; \citet{oviedo2026energy} build a bottom-up framework for frontier-scale inference and test-time scaling; \citet{you2025chatgpt} estimated the energy of a ChatGPT query; and EcoLogits estimates the impacts of calls to generative-AI services \citep{rince2025ecologits}. \emph{First-party disclosures} remain rare: Google published an in-situ fleet measurement with an explicit boundary \citep{elsworth2025measuring}, OpenAI stated an average per-query figure without a methodology \citep{altman2025gentle}, and Mistral AI published a life-cycle analysis of a large model \citep{mistral2025lca}. Journalists have combined measurements into accessible accounts \citep{odonnell2025math}, and analysts have begun to estimate the energy of coding agents from their own usage logs \citep{couch2026agents,hausfather2026agentic}.

At the other end of the scale, data-centre electricity is estimated from server shipments, facility data and scenarios \citep{masanet2020recalibrating,shehabi2024lbnl,iea2025energyai}. Linking these aggregates to the use people make of AI requires per-task values on a known boundary.

TokWatt does not produce new measurements. It is a structured, versioned record of published evidence, organized around the two variables that most often explain disagreement, the boundary and the functional unit, with a transparent set of rules for reconciling them. The functional unit is borrowed from life-cycle assessment \citep{iso14040}, the facility-overhead component follows the definition of power usage effectiveness \citep[PUE;][]{iso30134}, and the release follows the FAIR principles \citep{wilkinson2016fair} and ships a datasheet \citep{gebru2021datasheets}.

% ===========================================================================
\section{Data collection}
\label{sec:collection}

TokWatt covers the operational electricity of AI inference: the energy used to serve one functional unit, such as one answer, one image, one video clip, one task delegated to an agent, or one inference of a classic model. Training, the manufacture of hardware, networks and end-user devices are outside the scope of the energy values, but the boundary fields record whether a source included them.

An estimate is included when (i) the source reports its own measurement, model or disclosure for an identifiable functional unit, (ii) the value can be converted to watt-hours per functional unit, (iii) the method is described well enough to classify the boundary, or the absence of a description is itself recorded, (iv) the source is publicly accessible through a DOI, an arXiv identifier or a stable URL, and (v) it was published between 2023 and the release date. Secondary reports of another source's value are recorded only when they add information, are labelled as such, and never enter a reference set.

Sources were identified through web searches for per-query, per-image, per-video and per-agent-task energy figures and through citation chaining from the sources found, between 25 October 2026 and the release date. Version~\version{} favours widely cited first-hand sources. It is a curated selection, not a systematic review, and does not claim completeness.

For each estimate we record the value and unit exactly as published and their conversion to watt-hours (values per thousand inferences are divided by 1{,}000 and joules by 3{,}600), any interval with its type (interquartile range, range, standard deviation or scenario range), the model, provider, hosting platform and hardware, input and output tokens, co-reported CO$_2$e and water, the period the data describe, and a locator pointing to the table, figure or section from which the value was taken. Inconsistencies inside a source are documented in a notes field rather than resolved silently: for example, the text and a table of \citet{elsworth2025measuring} give different emissions and water values for the same narrow-boundary prompt, and one standard deviation in \citet{jegham2025hungry} is printed without its decimal point.

\verificationSentence{} New rows added in later releases go through the same check before they enter a reference set.

% ===========================================================================
\section{Data records}
\label{sec:records}

The release consists of ten UTF-8 CSV tables (Table~\ref{tab:files}), a data dictionary, a datasheet, a citation file and the scripts described in Section~\ref{sec:validation}.

\begin{table}[t]
  \centering
  \small
  \caption{Files in TokWatt version~\version.}
  \label{tab:files}
  \begin{tabularx}{\linewidth}{@{}l r X@{}}
    \toprule
    File & Rows & Content \\
    \midrule
    \file{estimates.csv} & \nEst & One row per published estimate: task, functional unit, model, provider, host, hardware, tokens, energy with interval, five boundary components, method, harmonization factor and harmonized energy, CO$_2$e and water where reported, locator, verification status \\
    \file{sources.csv} & \nSrcAll & Bibliographic record of every source (type, peer-review status, date, DOI or URL, access date) \\
    \file{tasks.csv} & \nTasks & Task taxonomy, with the rule that selects each task's reference set \\
    \file{task\_defaults.csv} & \nTasksRef & Derived reference value and range per task (generated) \\
    \file{token\_factors.csv} & \nTokenFactors & Energy per million input, output and cache-read tokens from two published analyses \citep{couch2026agents,gwittebolle2026claudecarbon} \\
    \file{macro.csv} & \nMacro & Data-centre electricity and power-demand estimates and projections \citep{iea2025energyai,gartner2026dc,shehabi2024lbnl} \\
    \file{grid\_carbon.csv} & \nGrid & Life-cycle carbon intensity of electricity for \nGridCountries{} countries and \nGridRegions{} regions, latest year from \gridYearMin{} to \gridYearMax{} \citep{owid2026energy} \\
    \file{equivalents.csv} & \nEquivalents & Everyday reference energies with their basis \\
    \file{boundary\_components.csv} & \nBoundaryComp & Boundary vocabulary \\
    \file{parameters.csv} & \nParams & Harmonization and calculator parameters with their sources \\
    \bottomrule
  \end{tabularx}
\end{table}

Tasks are grouped into five families. \emph{Text}: a short chat answer (about 100--500 input and 250--500 visible output tokens, no extended reasoning), a longer answer (about 1{,}000 tokens each way), a reasoning answer (including hidden reasoning tokens), a chat over a long document (about 10{,}000 input tokens), a very long input (about 100{,}000 tokens) and an AI-generated search answer. \emph{Agentic}: one task delegated to an agent, including every model call and tool step it triggers, one coding-agent session and one developer-day of coding agents. \emph{Image} and \emph{video}: one generated image and one short clip. \emph{Classic}: nine single-inference tasks such as text classification, object detection and summarization. Each task has a stated rule for its reference set (Section~\ref{sec:harmonization}).

Each estimate records five components (Table~\ref{tab:boundary}) with one of five states: \emph{yes} (counted), \emph{approx} (approximated, for example by a heuristic or an assumed factor), \emph{partial} (part of the component counted), \emph{no} (not counted) or \emph{unknown} (not stated by the source).

\begin{table}[t]
  \centering
  \small
  \caption{Boundary components recorded for every estimate.}
  \label{tab:boundary}
  \begin{tabularx}{\linewidth}{@{}l l X@{}}
    \toprule
    Component & Field & What it covers \\
    \midrule
    Accelerators & \file{incl\_accelerator} & Energy of the GPUs or TPUs that serve the request \\
    Host & \file{incl\_host} & Processors, memory and other server components around the accelerators \\
    Idle capacity & \file{incl\_idle} & Machines provisioned for reliability and peak load but idle at the time \\
    Facility overhead & \file{incl\_overhead\_pue} & Cooling, power conversion and other facility loads, usually via PUE \\
    Embodied & \file{incl\_embodied} & Manufacture of the hardware, amortized over its use \\
    \bottomrule
  \end{tabularx}
\end{table}

The \file{method} field takes one of seven values: \emph{measured} (energy metered while serving the workload), \emph{modeled} (computed from throughput, power and utilization assumptions), \emph{disclosed} (a figure stated by the operator without a method), \emph{estimated from tokens} (token counts multiplied by per-token factors), \emph{LCA} (life-cycle analysis), \emph{GPU~$\times$2} (measured GPU energy doubled to approximate the full system, as in the analysis by \citealp{odonnell2025math}), and \emph{secondary} (another source's value as reported elsewhere).

Version~\version{} holds \nEst{} estimates, \nEstEnergy{} of them with a point value, from \nSrcEst{} sources published between \yearMin{} and \yearMax{} and describing \nModels{} distinct models or systems (Table~\ref{tab:sources}). Most estimates are modeled (\nModeled), followed by measured (\nMeasured), measured GPU energy doubled to approximate the full system (\nGpuTwo) and token-based estimates (\nTokenBased); there is one disclosed value, one life-cycle analysis and one secondary report. Of these estimates, \nProprietary{} concern proprietary models and \nOpenWeights{} open-weight models. Intervals are available for \nInterval{} estimates (\pctInterval\%) and token counts for \nTokenRows, but only \nCOtwo{} estimates co-report CO$_2$e and \nWater{} co-report water. Peer-reviewed sources account for \nSrcPeer{} of the \nSrcEst. One modeling study contributes \nJegham{} of the \nEst{} rows \citep{jegham2025hungry}. The reference-set rules limit its weight for short chat answers (\nChatJegham{} of \refChatN{} rows), but it supplies \nReasonJegham{} of the \refReasonN{} rows behind the reasoning value, which therefore inherits that study's assumptions.

\begin{table}[t]
  \centering
  \footnotesize
  \caption{Sources that contribute estimates, ordered by number of estimates ($n$). All \nSrcAll{} sources, including those of the context tables, are listed in \file{sources.csv}.}
  \label{tab:sources}
  \input{generated/tab_sources}
\end{table}

% ===========================================================================
\section{Harmonization and reference values}
\label{sec:harmonization}

Estimates are harmonized to a \emph{server plus facility overhead} boundary: accelerators, host and facility overhead are included; idle capacity and embodied energy are not. Most sources either report this boundary or can be mapped onto it with one documented factor, whereas idle capacity is fully counted by a single estimate and embodied energy by none of the estimates that report a per-task value (Section~\ref{sec:results}).

Each estimate $E_i$ is multiplied by a factor $f_i$ set by its boundary flags:
\begin{equation}
  E^{\mathrm{h}}_i = f_i\,E_i, \qquad
  f_i =
  \begin{cases}
    F & \text{accelerators counted, host not counted,}\\
    \mathrm{PUE}_{\mathrm{ref}} & \text{host counted or approximated, overhead not counted,}\\
    1 & \text{otherwise.}
  \end{cases}
  \label{eq:harmonization}
\end{equation}
$F=\Facc$ is the ratio between Google's comprehensive per-prompt energy and its active-accelerator component measured in production: accelerators account for 58\% of the comprehensive value (0.14 of 0.24\Wh), so $F = 1/0.58$ \citep{elsworth2025measuring}. $\mathrm{PUE}_{\mathrm{ref}}=\PUEref$ is the median of the facility values used in the sources: Google's fleet average of 1.09 \citep{elsworth2025measuring} and the values of 1.12 (Microsoft Azure), 1.14 (Amazon Web Services) and 1.27 (DeepSeek) assumed by \citet{jegham2025hungry}. The last case of Eq.~\eqref{eq:harmonization} covers estimates already at or beyond the target boundary, estimates using the GPU-doubling heuristic (the doubling already stands in for the other components) and estimates whose boundary is not stated, which are flagged. Of the \nEstEnergy{} estimates with an energy value, \nFacc{} are scaled by $F$, \nFpue{} by $\mathrm{PUE}_{\mathrm{ref}}$ and \nFone{} are left unchanged. Both parameters are stored in \file{parameters.csv} and can be changed on the command line.

Pooling every estimate for a task would mix models of very different sizes and deployments of very different efficiency. Each task therefore has a stated rule that selects its reference set $R_t$, recorded in \file{tasks.csv}. For a short chat answer, the set contains first-hand estimates for the default model of a major consumer assistant in 2025--2026 or for a generic frontier-scale production deployment; for a reasoning answer, all reasoning-model rows at the medium prompt size plus generic estimates of test-time scaling; for video, clips of at least 4.5~seconds at 15 frames per second or more from models with at least five billion parameters or described by their source as higher quality. The reference value and range are
\begin{equation}
  \hat{E}_t = \operatorname*{median}_{i \in R_t} E^{\mathrm{h}}_i, \qquad
  \hat{E}^{\,\mathrm{low}}_t = \min_{i \in R_t} f_i L_i, \qquad
  \hat{E}^{\,\mathrm{high}}_t = \max_{i \in R_t} f_i U_i,
  \label{eq:reference}
\end{equation}
where $[L_i, U_i]$ is the interval reported for estimate $i$ (interquartile range, range, scenario range, or mean $\pm$ one standard deviation clipped at zero) and $L_i = U_i = E_i$ when no interval is reported. The median limits the influence of a single extreme estimate; the range is deliberately wide, as it should be when sources disagree. Table~\ref{tab:reference} lists the resulting values, which Section~\ref{sec:results} discusses.

\begin{table}[t]
  \centering
  \footnotesize
  \caption{Reference values per task, ordered by energy. Reference: median of the harmonized reference set; range: widest harmonized interval reported in the set (``--'' when no interval is reported); all-rows median: median of every harmonized estimate for the task. $^{\dagger}$Single source.}
  \label{tab:reference}
  \input{generated/tab_reference}
\end{table}

To connect per-task values with grid capacity, TokWatt uses a unit identity. If $N$ people each use $E$ watt-hours per day, the average power is $P = NE/(24~\mathrm{h})$. For $N = 10^{9}$,
\begin{equation}
  P~[\mathrm{GW}] = \frac{E~[\mathrm{Wh\,day^{-1}}]}{24}, \qquad
  W_{\mathrm{year}}~[\mathrm{TWh}] = 0.365\,E~[\mathrm{Wh\,day^{-1}}].
  \label{eq:billion}
\end{equation}
Every 24\Wh{} per person per day across a billion people is therefore an average draw of one gigawatt, about the output of a large power reactor.

% ===========================================================================
\section{Technical validation}
\label{sec:validation}

The script \file{validate.py} checks controlled vocabularies, referential integrity between all tables (including the source through which a secondary value was reported), the consistency of intervals with point values, every derived column and the contents of \file{task\_defaults.csv}, the exclusion of secondary reports from reference sets, and the recency of the grid data. Version~\version{} passes with no errors or warnings. Every value in this article that is derived from the database, and every table and figure, is generated from the released files by one script (\file{paper/analysis.py}) and can be regenerated.

Google reports two values for the same population of prompts: \googleFull\Wh{} with its comprehensive boundary and \googleNarrow\Wh{} with an ``existing approach'' that counts only active accelerators \citep{elsworth2025measuring}. Applying $F$ to the narrow value gives \googleHarm\Wh, \googleResidual{} times the comprehensive value. A residual gap is expected: the existing approach also samples highly utilized machines in the top 10\% most energy-efficient data centres, and boundary harmonization cannot correct for utilization. Harmonized accelerator-only measurements will therefore understate production energy whenever the measured system runs at a higher utilization than a production fleet.

\begin{figure}[t]
  \centering
  \includegraphics[width=\linewidth]{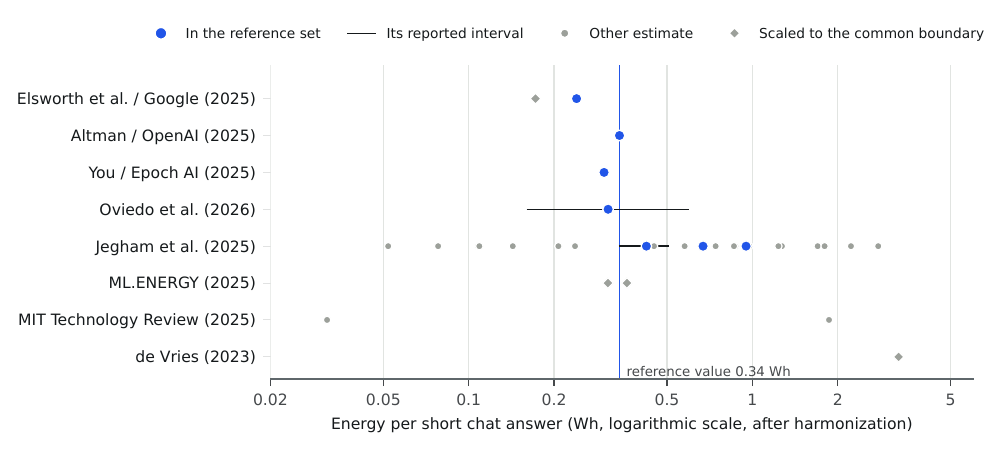}
  \caption{Published estimates of the energy of one short chat answer, after harmonization to the server-plus-facility boundary. Blue points form the reference set and the vertical line marks its median. Diamonds were scaled by Eq.~\eqref{eq:harmonization}: Google's accelerator-only value and the two ML.ENERGY values by $F=\Facc$ and the de~Vries estimate by $\mathrm{PUE}_{\mathrm{ref}}=\PUEref$. Logarithmic scale.}
  \label{fig:chat}
\end{figure}

The \chatAllN{} harmonized estimates for a short chat answer span a factor of \chatAllRatio{}, from \chatAllMin{} to \chatAllMax\Wh{} (Figure~\ref{fig:chat}). The spread narrows sharply once the functional unit and the deployment are matched. The five estimates that describe the default model of ChatGPT or Gemini, or a generic frontier-scale production deployment (one fleet measurement, one disclosure and three bottom-up models from different groups), lie between \clusterMin{} and \clusterMax\Wh, a factor of \clusterRatio. Adding the two other rows of the reference set, modeled responses of Claude~3.7 Sonnet and of GPT-5 with minimal reasoning, widens the span to a factor of \chatRefSpan. The extremes of the full set reflect model size and method rather than disagreement about the same system: the lowest value is an eight-billion-parameter open model \citep{odonnell2025math}, and the highest is the 2023 estimate derived from early analyst assumptions \citep{devries2023growing}.

Table~\ref{tab:sensitivity} recomputes the reference values of the six tasks offered in the calculator and of text classification under four alternatives: $F=2$ (the doubling heuristic), $\mathrm{PUE}_{\mathrm{ref}}$ at the lowest and highest values in its source set (1.09 and 1.27), and no harmonization at all. The reference values of \nInvariant{} of these \nSensTasks{} tasks do not change in any scenario, because the median of their reference set falls on an estimate that already reports the target boundary. The largest change is \sensMaxPct\%, for \sensChangedTasks, whose reference sets rest on server-level measurements without facility overhead. The uncertainty of the reference values is therefore dominated by the disagreement between sources, which the ranges carry, rather than by the harmonization parameters.

\begin{table}[t]
  \centering
  \small
  \caption{Reference values (Wh) under alternative harmonization parameters. Baseline: $F=\Facc$, $\mathrm{PUE}_{\mathrm{ref}}=\PUEref$. ``None'' applies no harmonization. Three significant figures are kept to make small changes visible.}
  \label{tab:sensitivity}
  \input{generated/tab_sensitivity}
\end{table}

% ===========================================================================
\section{Results}
\label{sec:results}

\begin{figure}[t]
  \centering
  \includegraphics[width=\linewidth]{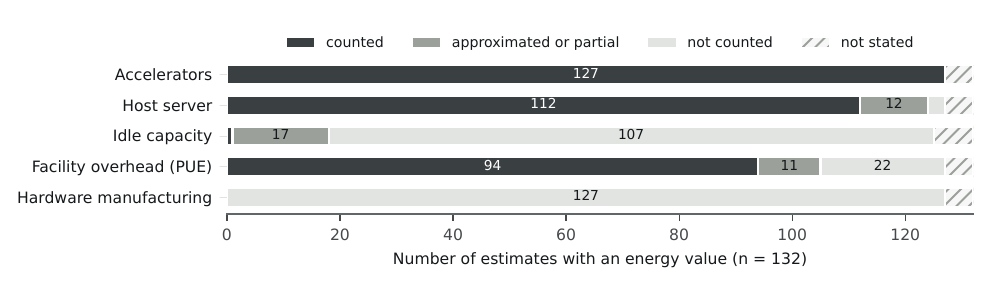}
  \caption{Which parts of the system published estimates count. Each bar covers the \nEstEnergy{} estimates that report an energy value.}
  \label{fig:boundary}
\end{figure}

Boundary reporting is uneven (Figure~\ref{fig:boundary}). Accelerator energy is counted in \bndAccYes{} of the \nEstEnergy{} estimates and host energy in \bndHostYes, but idle capacity is fully counted in only \bndIdleYes{} (approximated or partly counted in \bndIdleApprox), facility overhead in \bndPueYes{} (approximated in \bndPueApprox), and embodied energy in none of the estimates that report a per-task value; the one life-cycle analysis in the database, of \lcaModel, does not disclose energy per response \citep{mistral2025lca}. Only \nFullOperational{} estimates, from \nFullOperationalSrc{} sources, cover all four operational components, and \nAllUnknown{} state no boundary at all, among them the OpenAI disclosure \citep{altman2025gentle}.

Reference values span \oomWord{} orders of magnitude, from \refMin\Wh{} for a \minTaskLabel{} inference to \refMax\Wh{} for \maxTaskLabel{} (Figure~\ref{fig:ladder} and Table~\ref{tab:reference}). Relative to a short chat answer (\refChat\Wh), a reasoning answer uses about \ratioReason{} times as much energy, a chat over a long document \ratioLongdoc{} times, an image \ratioImage{} times, a short video clip \ratioVideo{} times and one task delegated to a coding agent \ratioAgent{} times; a short chat answer in turn uses about \ratioChatTextcls{} times as much as a classic text classification. For comparison, Google reported 0.3\Wh{} for a web search in 2009 \citep{google2009search}, the same order of magnitude as a short chat answer today. The ranges are wide where evidence is heterogeneous: the high-to-low ratio is \spanChat{} for a short chat answer, \spanImage{} for an image, \spanReason{} for a reasoning answer and \spanVideo{} for a video clip, whose reference set runs from \refVideoLo\Wh{} for a 13-billion-parameter model measured on a single H100 GPU \citep{delavande2025videoblog} to \refVideoHi\Wh{} for a newer version of CogVideoX \citep{odonnell2025math}. Of the \nTasksRef{} reference values, \nSingleSource{} rest on a single source and should be read as indicative.

Agentic use changes the unit of analysis. One prompt to a coding agent triggers many model calls that re-read a growing context, so its energy is set by the tokens processed rather than by the visible answer. The two analyses in the database derive energy from their authors' own coding-agent logs and per-token factors \citep{couch2026agents,hausfather2026agentic}: one delegated task is estimated at \refAgent\Wh{} (\refAgentLo--\refAgentHi), a session at \refSession\Wh{} (the midpoint of two estimates, \sessionLo{} and \sessionHi\Wh, that differ by a factor of about \sessionRatio) and a developer-day at \refDevday\Wh. These are the least certain values in the database. They describe two individuals' usage, rely on per-token factors derived from other estimates rather than measured, and depend on the energy of reading cached context, which no provider discloses and which drives most of the range: \citet{hausfather2026agentic} varies it between 1\% and 25\% of the energy of fresh input. The per-token factors released with TokWatt make the asymmetry explicit: \citet{couch2026agents} estimates \tfOut\Wh{} per million output tokens, \tfIn\Wh{} per million fresh input tokens and \tfCache\Wh{} per million cache reads.

Table~\ref{tab:profiles} applies the reference values to illustrative daily profiles and Eq.~\eqref{eq:billion}. The calculator's default day (twenty short answers, three reasoning answers, one long document, one agent task and two images) adds up to \exDayWh\Wh{} (range \exDayLo--\exDayHi), of which the single agent task accounts for \exAgentShare\%. Across a billion people this is an average draw of \exDayGW~GW, or \exDayTWh~TWh per year, about \exDayPctDC\% of the \gartnerTwhB~TWh projected for all data centres in 2026 \citep{gartner2026dc}. Chat alone is a small load at the volumes reported so far: one billion messages a day, the volume OpenAI reported for ChatGPT in December 2024 \citep{odonnell2025math}, correspond to about \chatgptMW~MW of continuous power at \refChat\Wh{} per answer. Heavy agent use is a different matter: twenty short answers and ten delegated agent tasks per person per day across a billion people would draw about \devGW~GW, close to the average power of all data centres projected for 2026 (\dcAvgGW~GW). These profiles illustrate scale and are not forecasts. They leave out training, idle capacity and embodied energy, and they ignore efficiency gains, which have been large: Google reports a 33-fold reduction in the energy of its median text prompt between May 2024 and May 2025 \citep{elsworth2025measuring}.

\begin{table}[t]
  \centering
  \footnotesize
  \caption{Illustrative daily profiles scaled to one billion users with Eq.~\eqref{eq:billion}. $E$ uses the reference values and the range the reference ranges. Power and energy are the average draw and annual consumption for $10^9$ users; the last column compares that energy with the \gartnerTwhB~TWh projected for all data centres worldwide in 2026 \citep{gartner2026dc}.}
  \label{tab:profiles}
  \input{generated/tab_profiles}
\end{table}

% ===========================================================================
\section{The TokWatt calculator}
\label{sec:app}

The calculator is a single static web page of about 160~kB with the database embedded, so it runs entirely in the browser without a server. A user counts a typical day of AI use by task, and the page returns the energy in watt-hours with its reference range, everyday equivalents, an electricity cost (by default \$\priceUSD{} per kWh, the 2025 average residential price forecast for the United States by \citealp{hodge2025eia}; editable), and CO$_2$e computed with the life-cycle carbon intensity of electricity of any of \nGridCountries{} countries or \nGridRegions{} regions \citep{owid2026energy}. A ``one billion people'' panel applies Eq.~\eqref{eq:billion} and compares the result with data-centre totals. The page also contains the energy ladder, an evidence explorer that lists every estimate with its boundary flags and source, downloads of the CSV tables, and a shareable image of the user's daily total. It is generated from the CSV files by \file{build\_app.py}, so each release of the data comes with a matching release of the calculator, which states the data version it uses.

% ===========================================================================
\section{Usage notes}
\label{sec:usage}

The tables can be read with any CSV reader; \file{estimates.csv} joins \file{sources.csv} on \file{source\_id} and \file{tasks.csv} on \file{task}. After any edit, the derived values, checks and calculator are rebuilt with the Python standard library only:
\par\noindent\begin{minipage}{\linewidth}\small
\begin{verbatim}
python scripts/build.py      # derived columns and task_defaults.csv
python scripts/validate.py   # checks
python scripts/build_app.py  # calculator
python scripts/build.py --factor-accel 2.0 --pue-ref 1.2   # sensitivity run
\end{verbatim}
\end{minipage}\par\smallskip
\noindent The last line shows how the harmonization parameters can be changed for a sensitivity run.

TokWatt is intended for comparing published estimates on a like-for-like basis, scenario analysis of AI electricity demand, teaching, calculators and dashboards that need sourced per-task values with uncertainty, and research on measurement practice. It is not intended for ranking named commercial products as if the values were audited measurements, for regulatory or compliance reporting, or for claims about the emissions of a specific data centre.

Releases follow semantic versioning and are published as tagged GitHub releases; studies should cite the version they used. Changes are listed in \file{CHANGELOG.md}.

A new estimate needs a first-hand public source with a DOI or stable URL, the functional unit, the value and unit exactly as published, the five boundary components and a locator. Contributions are made through pull requests that pass \file{validate.py}; the maintainer checks each new row against its source before it is marked as verified.

% ===========================================================================
\section{Limitations}
\label{sec:limitations}

The database is a curated selection of English-language sources that are visible on the web, not a systematic review. First-party disclosures are few: among the sources reviewed, only Google, OpenAI and Mistral AI publish per-query figures, and Mistral AI's describe emissions and water rather than energy. One modeling study supplies most rows.

Harmonization aligns boundaries; it does not correct for utilization, batching, hardware generation, quantization or serving software, which modeled estimates assume and benchmarks fix (Section~\ref{sec:validation}). The target boundary excludes idle capacity and embodied energy, so TokWatt values are lower than comprehensive footprints; in Google's measurement idle machines alone account for 10\% of the comprehensive per-prompt energy \citep{elsworth2025measuring}.

Of the \nTasksRef{} reference values, \nSingleSource{} rest on one source. The agentic values come from two individuals' logs and an undisclosed cache-read energy; video values come from open models at modest resolutions, and the classic tasks from one study run without batching \citep{luccioni2024power}. No provider has published measurements for frontier closed video or agent systems.

Efficiency improves quickly, as Google's 33-fold reduction in a year shows, while workloads grow longer through reasoning and agents. Every estimate carries its data period, and the reference values should be read as describing \yearMin--\yearMax{} systems.

% ===========================================================================
\section{Conclusion}
\label{sec:conclusion}

Much of the disagreement about the energy of AI is a disagreement about what is counted and what is being done. Recording both for every published estimate makes the evidence comparable: once the boundary and the functional unit are matched, independent estimates of a default chat answer agree within a factor of \clusterRatio, and the remaining spread across tasks, \oomWord{} orders of magnitude from a classic inference to a day of coding agents, becomes the informative signal. The shift from prompts to agents is the main reason per-person use may matter for the grid, and it is also where the evidence is thinnest. TokWatt is designed to be extended as providers disclose more and as measurements of agents, video and reasoning systems appear; contributions of new first-hand estimates are welcome.

% ===========================================================================
\section*{Data availability}
The TokWatt database (version~\version) is available at \repoURL{} under the CC~BY~4.0 licence, with each version published as a tagged GitHub release \citep{bahi2026tokwatt}. Grid carbon-intensity values are from Our World in Data and Ember, also under CC~BY~4.0 \citep{owid2026energy}.

\section*{Code availability}
The build, validation and calculator code is available in the same repository under the MIT licence. The script \file{paper/analysis.py} regenerates every database-derived number, table and figure in this article from the released data.

\section*{Use of AI tools}
An AI assistant (Claude, Anthropic) was used to help search for sources, transcribe values, write code and edit the text. The author designed the study, checked every value and every reference against its source, and takes full responsibility for the content of this article and of the database.

\section*{Acknowledgements}
The author thanks the researchers, companies and journalists whose published measurements, models and disclosures make this database possible.

\section*{Competing interests}
The author declares no competing interests.

\bibliographystyle{plainnat}
\bibliography{references}

\end{document}

%% file: generated/stats.tex
\newcommand{\nEst}{134}
\newcommand{\nEstEnergy}{132}
\newcommand{\nSrcEst}{13}
\newcommand{\nSrcAll}{22}
\newcommand{\nTasks}{20}
\newcommand{\nTasksRef}{19}

\newcommand{\nModels}{57}
\newcommand{\nGrid}{220}
\newcommand{\nGridCountries}{212}
\newcommand{\nGridRegions}{8}
\newcommand{\gridYearMin}{2022}
\newcommand{\gridYearMax}{2025}
\newcommand{\nMacro}{31}
\newcommand{\nTokenFactors}{6}
\newcommand{\nEquivalents}{10}
\newcommand{\nMeasured}{22}
\newcommand{\nModeled}{98}

\newcommand{\nTokenBased}{5}

\newcommand{\nGpuTwo}{6}

\newcommand{\nProprietary}{65}
\newcommand{\nOpenWeights}{69}
\newcommand{\nSrcPeer}{4}
\newcommand{\yearMin}{2023}
\newcommand{\yearMax}{2026}
\newcommand{\nInterval}{103}
\newcommand{\pctInterval}{78}
\newcommand{\nTokenRows}{98}
\newcommand{\nCOtwo}{3}
\newcommand{\nWater}{4}

\newcommand{\nFacc}{3}
\newcommand{\nFpue}{19}
\newcommand{\nFone}{110}
\newcommand{\Facc}{1.72}
\newcommand{\PUEref}{1.13}

\newcommand{\priceUSD}{0.168}
\newcommand{\bndAccYes}{127}

\newcommand{\bndHostYes}{112}

\newcommand{\bndIdleYes}{1}
\newcommand{\bndIdleApprox}{17}

\newcommand{\bndPueYes}{94}
\newcommand{\bndPueApprox}{11}

\newcommand{\nFullOperational}{7}
\newcommand{\nFullOperationalSrc}{2}
\newcommand{\nAllUnknown}{5}
\newcommand{\lcaModel}{Mistral Large 2}

\newcommand{\verificationSentence}{The author checked every row against the table, figure or section cited as its locator; all 134 rows carry the status \texttt{human\_verified} in the \texttt{verification} field.}
\newcommand{\googleNarrow}{0.10}
\newcommand{\googleFull}{0.24}
\newcommand{\googleHarm}{0.17}
\newcommand{\googleResidual}{0.72}

\newcommand{\clusterMin}{0.24}
\newcommand{\clusterMax}{0.42}
\newcommand{\clusterRatio}{1.8}
\newcommand{\chatRefSpan}{4.0}

\newcommand{\chatAllN}{33}
\newcommand{\chatAllMin}{0.032}
\newcommand{\chatAllMax}{3.3}
\newcommand{\chatAllRatio}{100}

\newcommand{\refChat}{0.34}
\newcommand{\refChatLo}{0.16}
\newcommand{\refChatHi}{0.99}
\newcommand{\refChatN}{7}

\newcommand{\refReason}{4.7}

\newcommand{\refReasonN}{10}

\newcommand{\refAgent}{150}
\newcommand{\refAgentLo}{60}
\newcommand{\refAgentHi}{290}

\newcommand{\refSession}{321}

\newcommand{\refDevday}{2,150}

\newcommand{\refVideo}{124}
\newcommand{\refVideoLo}{4.3}
\newcommand{\refVideoHi}{944}

\newcommand{\ratioReason}{14}
\newcommand{\ratioLongdoc}{8.5}
\newcommand{\ratioImage}{3.6}
\newcommand{\ratioVideo}{360}
\newcommand{\ratioAgent}{440}

\newcommand{\ratioChatTextcls}{150}

\newcommand{\oomWord}{six}
\newcommand{\minTaskLabel}{text classification}
\newcommand{\maxTaskLabel}{one developer-day of coding agents}
\newcommand{\refMin}{0.0023}
\newcommand{\refMax}{2,150}
\newcommand{\nSingleSource}{12}
\newcommand{\nBoundaryComp}{5}
\newcommand{\nParams}{5}
\newcommand{\mitLo}{0.03}
\newcommand{\mitHi}{1.9}
\newcommand{\nJegham}{91}
\newcommand{\nChatJegham}{3}
\newcommand{\nReasonJegham}{9}
\newcommand{\spanReason}{30}
\newcommand{\spanImage}{20}
\newcommand{\spanVideo}{220}
\newcommand{\spanChat}{6.2}
\newcommand{\tfIn}{390}
\newcommand{\tfOut}{1,950}
\newcommand{\tfCache}{39}

\newcommand{\sessionLo}{41}
\newcommand{\sessionHi}{600}
\newcommand{\sessionRatio}{15}
\newcommand{\ieaTwhA}{415}
\newcommand{\ieaShare}{1.5}
\newcommand{\ieaTwhB}{945}
\newcommand{\gartnerTwhA}{447}
\newcommand{\gartnerTwhB}{565}

\newcommand{\gartnerAIShare}{31}

\newcommand{\gartnerTwhC}{1,200}
\newcommand{\lbnlTwh}{176}
\newcommand{\lbnlShare}{4.4}
\newcommand{\lbnlTwhLo}{325}
\newcommand{\lbnlTwhHi}{580}
\newcommand{\lbnlShareLo}{6.7}
\newcommand{\lbnlShareHi}{12}
\newcommand{\dcAvgGW}{64.5}

\newcommand{\chatgptMW}{14}

\newcommand{\exDayWh}{176}
\newcommand{\exDayLo}{70}
\newcommand{\exDayHi}{444}
\newcommand{\exDayGW}{7.3}
\newcommand{\exDayTWh}{64}
\newcommand{\exDayPctDC}{11}
\newcommand{\exAgentShare}{85}
\newcommand{\devGW}{63}

\newcommand{\sensMaxPct}{12}

\newcommand{\nInvariant}{5}
\newcommand{\sensChangedTasks}{video generation and text classification}

\newcommand{\nSensTasks}{7}

%% file: figures/fig_overview.tex
% Figure 1: overview of the TokWatt database (TikZ, numbers from generated/stats.tex)
\begin{tikzpicture}[
  font=\sffamily,
  x=1cm, y=1cm,
  twcard/.style={draw=twLine, line width=0.6pt, fill=twWash, rounded corners=7pt, minimum width=3.55cm, minimum height=4.35cm},
  twout/.style={draw=twLine, line width=0.6pt, fill=white, rounded corners=6pt, minimum width=4.95cm, minimum height=1.05cm},
  twflow/.style={-{Stealth[length=6pt,width=5pt]}, line width=1.1pt, draw=twBlue!70},
  twnum/.style={text=twNavy, font=\sffamily\bfseries\Large},
  twlab/.style={text=twNavy, font=\sffamily\bfseries\footnotesize, align=center},
  twsub/.style={text=twGrey, font=\sffamily\scriptsize, align=center, text width=3.2cm, execute at begin node={\hyphenpenalty=10000\exhyphenpenalty=10000}},
  stepno/.style={circle, fill=twBlue, text=white, font=\sffamily\bfseries\scriptsize, inner sep=1.4pt, minimum size=12pt},
]
% ---------------------------------------------------------------- stage cards
\foreach \i/\x in {1/0, 2/4.15, 3/8.30, 4/12.45} {
  \node[twcard] (c\i) at (\x, 0) {};
  \node[stepno] at ([xshift=12pt, yshift=-12pt]c\i.north west) {\i};
}
\foreach \a/\b in {1/2, 2/3, 3/4} \draw[twflow] (c\a.east) -- (c\b.west);

% icon 1: a stack of documents
\begin{scope}[shift={(0, 1.05)}]
  \foreach \d/\c in {0.16/twBlue!25, 0.08/twBlue!45, 0/white} {
    \fill[\c, draw=twBlue, line width=0.7pt, rounded corners=1.5pt] (-0.42+\d, -0.52+\d) -- (0.22+\d, -0.52+\d) -- (0.22+\d, 0.30+\d) -- (0.04+\d, 0.48+\d) -- (-0.42+\d, 0.48+\d) -- cycle;
  }
  \draw[twBlue, line width=0.7pt] (0.04, 0.48) -- (0.04, 0.30) -- (0.22, 0.30);
  \foreach \yy in {0.22, 0.06, -0.10, -0.26} \draw[twBlue!70, line width=0.8pt, line cap=round] (-0.30, \yy) -- (0.10, \yy);
\end{scope}
\node[twnum] at (0, 0.05) {\nSrcEst};
\node[twlab] at (0, -0.42) {published sources};
\node[twsub] at (0, -1.25) {papers, company disclosures, benchmarks and journalism (\yearMin--\yearMax)};

% icon 2: a database cylinder
\begin{scope}[shift={(4.15, 1.05)}]
  \fill[twBlue!15, draw=twBlue, line width=0.7pt] (-0.42, -0.42) -- (-0.42, 0.34) arc[start angle=180, end angle=360, x radius=0.42, y radius=0.13] -- (0.42, -0.42) arc[start angle=0, end angle=-180, x radius=0.42, y radius=0.13];
  \fill[white, draw=twBlue, line width=0.7pt] (0, 0.34) ellipse[x radius=0.42, y radius=0.13];
  \foreach \yy in {0.08, -0.17} \draw[twBlue, line width=0.7pt] (-0.42, \yy) arc[start angle=180, end angle=360, x radius=0.42, y radius=0.13];
\end{scope}
\node[twnum] at (4.15, 0.05) {\nEst};
\node[twlab] at (4.15, -0.42) {estimates};
\node[twsub] at (4.15, -1.25) {\nTasks{} tasks, \nModels{} models or systems; value, interval and locator in the source};

% icon 3: nested boundary (chip, server, facility)
\begin{scope}[shift={(8.30, 1.05)}]
  \draw[twBlue, line width=0.7pt, fill=twBlue!8, rounded corners=1.5pt] (-0.62, -0.46) rectangle (0.62, 0.30);
  \draw[twBlue, line width=0.7pt, fill=twWash] (-0.68, 0.30) -- (0, 0.56) -- (0.68, 0.30) -- cycle;
  \draw[twBlue, line width=0.7pt, fill=white, rounded corners=1.5pt] (-0.40, -0.36) rectangle (0.40, 0.18);
  \fill[twBlue] (-0.13, -0.22) rectangle (0.13, 0.04);
  \foreach \xx in {-0.08, 0, 0.08} {\draw[twBlue, line width=0.5pt] (\xx, 0.04) -- (\xx, 0.10); \draw[twBlue, line width=0.5pt] (\xx, -0.22) -- (\xx, -0.28);}
\end{scope}
\node[twnum] at (8.30, 0.05) {5};
\node[twlab] at (8.30, -0.42) {boundary flags};
\begin{scope}[shift={(8.30, -0.92)}]
  \foreach \k/\t in {0/chip, 1/host, 2/idle, 3/PUE, 4/embodied} {
    \pgfmathsetmacro{\xx}{-1.32 + 0.66*\k}
    \fill[twBlue] (\xx, 0.12) circle (3.2pt);
    \node[text=twGrey, font=\sffamily\tiny] at (\xx, -0.13) {\t};
  }
  \node[twsub] at (0, -0.78) {harmonized to server + facility: $\times$\Facc{} chip-only, $\times$\PUEref{} server-only};
\end{scope}

% icon 4: a small energy ladder on a log axis
\begin{scope}[shift={(12.45, 1.05)}]
  \draw[twLine, line width=0.6pt] (-0.62, -0.46) -- (0.62, -0.46);
  \foreach \k/\a/\b/\m in {0/-0.60/-0.32/-0.47, 1/-0.36/-0.04/-0.21, 2/-0.14/0.24/0.03, 3/0.10/0.44/0.27, 4/0.30/0.62/0.44} {
    \pgfmathsetmacro{\yy}{-0.34 + 0.19*\k}
    \draw[twBlue!30, line width=3.2pt, line cap=round] (\a, \yy) -- (\b, \yy);
    \fill[twBlue] (\m, \yy) circle (2.4pt);
  }
\end{scope}
\node[twnum] at (12.45, 0.05) {\nTasksRef};
\node[twlab] at (12.45, -0.42) {reference values};
\node[twsub] at (12.45, -1.25) {median and range per task, from \refMin\,Wh to \refMax\,Wh};

% ---------------------------------------------------------------- outputs
\node[font=\sffamily\bfseries\scriptsize, text=twGrey] at (6.225, -2.72) {OPEN OUTPUTS};
\draw[twLine, line width=0.5pt] (-1.775, -2.72) -- (5.05, -2.72);
\draw[twLine, line width=0.5pt] (7.40, -2.72) -- (14.225, -2.72);
\foreach \i/\x in {1/0.70, 2/6.225, 3/11.75} \node[twout] (o\i) at (\x, -3.45) {};

% output 1: CSV tables
\begin{scope}[shift={(-1.15, -3.45)}]
  \draw[twBlue, line width=0.7pt, fill=white, rounded corners=1pt] (-0.30, -0.24) rectangle (0.30, 0.24);
  \fill[twBlue] (-0.30, 0.10) rectangle (0.30, 0.24);
  \draw[twBlue, line width=0.5pt] (-0.30, -0.07) -- (0.30, -0.07) (-0.10, -0.24) -- (-0.10, 0.10) (0.10, -0.24) -- (0.10, 0.10);
\end{scope}
\node[anchor=west, align=left, text=twNavy, font=\sffamily\footnotesize] at (-0.65, -3.45) {\textbf{\nEst{} rows, 10 CSV tables}\\[-1pt]{\scriptsize\color{twGrey}CC BY 4.0, rebuilt by open scripts}};

% output 2: browser calculator
\begin{scope}[shift={(4.37, -3.45)}]
  \draw[twBlue, line width=0.7pt, fill=white, rounded corners=1.5pt] (-0.32, -0.25) rectangle (0.32, 0.25);
  \fill[twBlue, rounded corners=1.5pt] (-0.32, 0.12) rectangle (0.32, 0.25);
  \fill[white] (-0.24, 0.185) circle (0.9pt) (-0.17, 0.185) circle (0.9pt);
  \fill[twBlue!35] (-0.22, -0.17) rectangle (-0.10, -0.02);
  \fill[twBlue!60] (-0.06, -0.17) rectangle (0.06, 0.05);
  \fill[twBlue] (0.10, -0.17) rectangle (0.22, 0.08);
\end{scope}
\node[anchor=west, align=left, text=twNavy, font=\sffamily\footnotesize] at (4.86, -3.45) {\textbf{Web calculator}\\[-1pt]{\scriptsize\color{twGrey}Wh, cost and CO\textsubscript{2}e of an AI day}};

% output 3: a lightning bolt for the billion-user rule
\begin{scope}[shift={(9.88, -3.45)}]
  \fill[twBlue!12] (0, 0) circle (0.30);
  \fill[twBlue] (0.05, 0.25) -- (-0.14, -0.03) -- (-0.01, -0.03) -- (-0.06, -0.25) -- (0.14, 0.05) -- (0.01, 0.05) -- cycle;
\end{scope}
\node[anchor=west, align=left, text=twNavy, font=\sffamily\footnotesize] at (10.37, -3.45) {\textbf{\textit{E} Wh a day, 1 billion users}\\[-1pt]{\scriptsize\color{twGrey}= \textit{E}\,/\,24 GW of average demand}};
\end{tikzpicture}

%% file: generated/tab_sources.tex
% generated by paper/analysis.py -- do not edit
\setlength{\tabcolsep}{4.5pt}
\begin{tabularx}{\linewidth}{@{}>{\raggedright\arraybackslash}X l l >{\raggedright\arraybackslash}X l r@{}}
\toprule
Source & Type & Peer review & Method & Task families & $n$ \\
\midrule
\citet{jegham2025hungry} & Preprint & no & modeled & text & 91 \\
\citet{luccioni2024power} & Conference paper & yes & measured & image, classic & 11 \\
\citet{delavande2025videoblog} & Benchmark report & no & measured & video & 7 \\
\citet{odonnell2025math} & Journalism & no & GPU $\times$2 & text, image, video & 6 \\
\citet{hausfather2026agentic} & Analyst estimate & no & secondary, token-based & agentic & 4 \\
\citet{you2025chatgpt} & Research note & no & modeled & text & 3 \\
\citet{devries2023growing} & Commentary & yes & modeled & text & 2 \\
\citet{elsworth2025measuring} & Company study & no & measured & text & 2 \\
\citet{chung2025mlenergy} & Conference paper & yes & measured & text & 2 \\
\citet{couch2026agents} & Analyst estimate & no & token-based & agentic & 2 \\
\citet{oviedo2026energy} & Journal article & yes & modeled & text & 2 \\
\citet{altman2025gentle} & Company statement & no & disclosed & text & 1 \\
\citet{mistral2025lca} & Company LCA & external & LCA & text & 1 \\
\midrule
Total & & & & & 134 \\
\bottomrule
\end{tabularx}

%% file: generated/tab_reference.tex
% generated by paper/analysis.py -- do not edit
\setlength{\tabcolsep}{4.5pt}
\begin{tabular}{@{}l r r r r r r r@{}}
\toprule
& Reference & Range & High/ & Median, & \multicolumn{2}{c}{Estimates} & \\
\cmidrule(lr){6-7}
Task & (Wh) & (Wh) & low & all rows (Wh) & ref.\ set & all & Sources \\
\midrule
Text classification$^{\dagger}$ & 0.0023 & 0.0011--0.0034 & 3.0 & 0.0023 & 1 & 1 & 1 \\
Extractive question answering$^{\dagger}$ & 0.0034 & 0.0023--0.0045 & 2.0 & 0.0034 & 1 & 1 & 1 \\
Masked language modelling$^{\dagger}$ & 0.0034 & 0.0023--0.0045 & 2.0 & 0.0034 & 1 & 1 & 1 \\
Token classification$^{\dagger}$ & 0.0045 & 0.0023--0.0068 & 3.0 & 0.0045 & 1 & 1 & 1 \\
Image classification$^{\dagger}$ & 0.0079 & 0.0068--0.0090 & 1.3 & 0.0079 & 1 & 1 & 1 \\
Object detection$^{\dagger}$ & 0.043 & 0.020--0.066 & 3.2 & 0.043 & 1 & 1 & 1 \\
Short text generation (10 tokens)$^{\dagger}$ & 0.053 & 0.019--0.087 & 4.5 & 0.053 & 1 & 1 & 1 \\
Summarization$^{\dagger}$ & 0.055 & 0.044--0.067 & 1.5 & 0.055 & 1 & 1 & 1 \\
Image captioning$^{\dagger}$ & 0.071 & 0.049--0.094 & 1.9 & 0.071 & 1 & 1 & 1 \\
Short chat answer & 0.34 & 0.16--0.99 & 6.2 & 0.45 & 7 & 33 & 5 \\
Generate one image & 1.2 & 0.63--13 & 20 & 1.4 & 3 & 4 & 2 \\
Longer chat answer$^{\dagger}$ & 2.3 & 0.97--3.2 & 3.3 & 2.3 & 3 & 23 & 1 \\
Chat over a long document & 2.9 & 2.5--6.0 & 2.4 & 3.4 & 3 & 20 & 2 \\
Reasoning answer & 4.7 & 1.1--34 & 30 & 4.3 & 10 & 27 & 2 \\
Very long input$^{\dagger}$ & 40 & -- & -- & 40 & 1 & 1 & 1 \\
Generate one short video & 124 & 4.3--944 & 220 & 29 & 3 & 9 & 2 \\
One task delegated to an AI agent$^{\dagger}$ & 150 & 60--290 & 4.8 & 150 & 1 & 1 & 1 \\
One coding-agent session & 321 & 41--1,200 & 29 & 459 & 2 & 3 & 2 \\
One developer-day of coding agents & 2,150 & 1,200--5,900 & 4.9 & 2,150 & 2 & 2 & 2 \\
\bottomrule
\end{tabular}

%% file: generated/tab_sensitivity.tex
% generated by paper/analysis.py -- do not edit
\begin{tabular}{@{}l r r r r r@{}}
\toprule
Task & Baseline & $F=2$ & PUE 1.09 & PUE 1.27 & None \\
\midrule
Short chat answer & 0.340 & 0.340 & 0.340 & 0.340 & 0.340 \\
Reasoning answer & 4.74 & 4.74 & 4.74 & 4.74 & 4.74 \\
Chat over a long document & 2.88 & 2.88 & 2.88 & 2.88 & 2.88 \\
One task delegated to an AI agent & 150 & 150 & 150 & 150 & 150 \\
Generate one image & 1.22 & 1.22 & 1.22 & 1.22 & 1.22 \\
Generate one short video & 124 & 124 & 119 & 139 & 109 \\
Text classification & 0.00226 & 0.00226 & 0.00218 & 0.00254 & 0.00200 \\
\bottomrule
\end{tabular}

%% file: generated/tab_profiles.tex
% generated by paper/analysis.py -- do not edit
\begin{tabularx}{\linewidth}{@{}>{\raggedright\arraybackslash}X r r r r r@{}}
\toprule
Profile and tasks per user per day & $E$ & Range & Power & Energy & Share of 2026 \\
& (Wh/day) & (Wh/day) & (GW) & (TWh/yr) & DC use (\%) \\
\midrule
\textbf{Occasional chat}\newline 5 chat & 1.7 & 0.80--5.0 & 0.071 & 0.62 & 0.11 \\
\textbf{Daily chat}\newline 20 chat, 3 reasoning & 21 & 6.6--122 & 0.88 & 7.7 & 1.4 \\
\textbf{Calculator example day}\newline 20 chat, 3 reasoning, 1 long doc., 1 agent, 2 image & 176 & 70--444 & 7.3 & 64 & 11 \\
\textbf{Knowledge work}\newline 30 chat, 10 reasoning, 5 long doc. & 72 & 28--400 & 3.0 & 26 & 4.7 \\
\textbf{Image and video creation}\newline 20 chat, 20 image, 2 video & 279 & 24--2,170 & 12 & 100 & 18 \\
\textbf{Agent-assisted development}\newline 20 chat, 10 agent & 1,510 & 603--2,920 & 63 & 550 & 97 \\
\bottomrule
\end{tabularx}